\documentclass[twocolumn]{autartArXiv}   

\usepackage{algorithm}
\usepackage{algorithmic}
\usepackage{amssymb}
\usepackage{url}
\usepackage{color}
\usepackage{graphicx}          

\newcommand{\Fi}{M}

\newcommand{\ta}{\theta}
\newcommand{\bydef}{\stackrel{\Delta}{=}}

\newcommand{\be}{\begin{equation}}
\newcommand{\ee}{\end{equation}}
\newcommand{\beqna}{\begin{eqnarray}}
\newcommand{\eeqna}{\end{eqnarray}}
\newcommand{\beqnd}{\begin{eqnarray*}}
\newcommand{\eeqnd}{\end{eqnarray*}}

\theoremstyle{plain} \newtheorem{definition}{Definition}  
\theoremstyle{definition} \newtheorem{assumption}{Assumption}
\theoremstyle{plain} \newtheorem{theorem}{Theorem} 
\theoremstyle{definition} \newtheorem{notation}{Notation}
\theoremstyle{definition} \newtheorem{remark}{Remark}
\theoremstyle{definition} \newtheorem{example}{Example} \theoremstyle{definition}  

\begin{document}

\begin{frontmatter}
\runtitle{Input Design for Nonlinear Systems}

\title{D-Optimal Input Design for Nonlinear FIR-type Systems: \\A
Dispersion-based Approach}

\author[Brussel]{Alexander De Cock}\ead{adecock@vub.ac.be},    
\author[Louvain]{Michel Gevers}\ead{Michel.Gevers@uclouvain.be},               
\author[Brussel]{Johan Schoukens}\ead{Johan.Schoukens@vub.ac.be}  

\address[Brussel]{ELEC, Vrije Universiteit Brussel, 1050 Brussel, Belgium}    
\address[Louvain]{ICTEAM, Louvain University, B1348 Louvain la Neuve, Belgium}


\begin{abstract}                      
Optimal input design is an important step of the identification process in order to reduce the model variance. In this work a D-optimal input design method for finite-impulse-response-type nonlinear systems is presented. The optimization of the determinant of the Fisher information matrix is expressed as a convex optimization problem. This problem is then solved using a dispersion-based optimization scheme, which is easy to implement and converges monotonically to the optimal solution. Without constraints, the optimal design cannot be realized as a time sequence. By imposing that the design should lie in the subspace described by a symmetric and non-overlapping set, a realizable design is found. A graph-based method is used in order to find a time sequence that realizes this optimal constrained design. These methods are illustrated on a numerical example of which the results are thoroughly discussed. Additionally the computational speed of the algorithm is compared with the general convex optimizer cvx.
\end{abstract}

\end{frontmatter}

\section{Introduction}
The quality of identified models  depends to a large extent on the experimental conditions under which the measurement data for these models are obtained.  Therefore, experiment design is an important step in the identification process. This was recognized in the early days of the development of system identification theory, where  significant attention was paid to the design of optimal experiments for the parametric  identification of linear time-invariant systems. The focus was on the design of optimal inputs  that maximize some scalar function of the Fisher information matrix  under a constraint on the power of the input signal \cite{Federov1972,Goodwin&Payne1977}. The motivation is that if the parameter estimator is asymptotically efficient, then its covariance matrix converges to the inverse of the Fisher information matrix. 
\\ For linear  stationary  systems operating in open loop, the Fisher Information matrix is an affine function of the input spectrum. Therefore, the optimization is  performed by first computing the optimal input spectrum, and then constructing an input signal for the experiment  that realizes this optimal input spectrum  \cite{Federov1972,Goodwin&Payne1977}. 
\\ The development of identification for control, and more generally of application-oriented identification \cite{Hjalmarsson:09},  together with the advent of powerful semi-definite programming tools, gave rise  to novel optimal experiment design techniques for linear time-invariant systems.  Experiment design methods were developed for closed-loop identification with a fixed controller \cite{Jansson&Hjalmarsson:05} as well as for closed-loop identification with a ``to be designed controller'' \cite{Hildebrand&Gevers&Solari:15}, and for an ever wider range of possible design criteria and  constraints. In addition, the dual problem  of least costly identification was addressed, where the design aims at minimizing the cost of the identification experiment (say, in terms of the input signal energy used) subject to a constraint on the achieved model quality \cite{Bombois&Scorletti&Gevers&Vandenhof&Hildebrand:06}. A survey of these results  can be found in \cite{Gevers&Bombois&Hildebrand&Solari:11}.
\\ For linear time-invariant systems, the solution of these optimal experiment design problems  consists of first expressing the criterion and the constraints in terms of a finite parametrization of the input spectrum (in the case of open loop design) or of the joint input-output spectrum (in the case of closed loop design), and reducing the problem into a convex optimization problem under Linear Matrix Inequality (LMI) constraints over these parameters. This yields an optimal spectrum. The second step then consists of constructing a stationary signal that has this desired spectrum.
A few results have also been developed where the optimization is performed directly with respect to the input samples of the signal in time-domain; see e.g. \cite{Cooley&Lee:01}.
\\ Optimal experiment design for nonlinear systems is considerably more difficult, because the criterion is typically non-convex,  making global optimization more difficult, if not impossible. The main difficulty is  that the Fisher information matrix of the experiment is not only dependent on the second order moments of the input but also on higher order moments \cite{Hjalmarsson2007}. Thus, optimal input design (OID)  results  for nonlinear systems have so far been obtained only for simple  design criteria in the form of scalar functions of the information matrix, and by restricting both the class of systems and the class of input signals. 
\\ One subclass of nonlinear systems for which a number of OID results have been obtained is the class of nonlinear finite-impulse-response-type (FIR-type) systems. Constructing an OID  for an example system out of this class was performed for Gaussian signals in \cite{MichelGevers2012} and for deterministic signals taking only a finite number of possible values 
in \cite{DeCock2013}. The restriction to such multilevel excitation  signals  is actually common to almost all recent results on OID for classes of nonlinear systems. 
In \cite{Larsson2010} a solution is proposed for nonlinear FIR systems using a probabilistic parametrization of the multilevel input signals, while in \cite{Forgione2014} these results are extended to systems with fading memory using deterministic multilevel input signals which are then characterized by the relative frequency of each possible subsequence. A common feature of these results is the adoption of a linear parametrization of the probability distributions, respectively the relative frequencies, of the input signals, leading to a convex formulation of the optimization criterion. 
\\ The difficult step in all OID methods  for classes of nonlinear systems proposed so far is to go from the optimal distributions (or optimal relative frequencies) to the generation of a realizable input sequence. A solution based on graph-theoretical properties has recently been proposed in \cite{Patricio_Automatica}: realizable input sequences are obtained by first computing the elementary cycles of the associated graph, which define a convex polyhedron. 
As we shall see, a significant part of the present paper is devoted to this problem of signal generation; it is also based on properties of the associated graph. By imposing additional symmetry constraints on the admissible subsequences,  we propose a suboptimal solution that is computationally cheaper than the solution based on elementary cycles described in \cite{Patricio_Automatica}. Additionally, we show that the optimization time for our solution compares favorably with that of a general purpose convex optimizer cvx.
\\ Methods considering a wider class of nonlinear systems also exist. For example in \cite{Gopaluni2011} a particle filter approach for the general class of nonlinear systems is presented, while  \cite{Vincent2010} presents an OID for a block structured nonlinear model consisting of linear dynamic and nonlinear static blocks. However, unlike the design methods for  nonlinear FIR-type systems, these methods  do not  result in a convex optimization problem and can therefore not guarantee convergence to a global optimum. 
\\ To conclude this general introduction, let us mention that OID for nonlinear systems is particular interesting in fields where the cost of single experiment is very high. Examples of practically applied OID for nonlinear systems can be found in the field of (bio)chemistry \cite{Telen2014}, cellular biology \cite{Dinh2014}  and medicine \cite{Galvanin2009}. An extensive overview of the current state-of-the-art can be found in \cite{Franceschini2008}.
\emph{Contribution and relation with other work}\\
This work is restricted to nonlinear FIR-type systems of memory length $n$, whose inputs are \emph{deterministic} input sequences of  length $n$, of which the sample values $u(t)$ can only take a finite set of possible values: $\{ u_1 , . . . , u_A \}$. Therefore each output sample of the system depends on a subsequence $(u(t),u(t-1),...u(t-n+1))$ of the input sequence, and  the total number of different subsequences that can be presented to the system is limited to a finite set of $A^n$ possible sequences. The Fisher information produced by a given input sequence is therefore completely determined by the subsequences it contains. For each input sequence of length $N$ we can define a corresponding \emph{frequency vector} that indicates how many times each subsequence is present in the sequence. It will then be shown that the Fisher information matrix  can be written as a convex combination of $A^n$ elementary Fisher information matrices,  where the coefficients are the relative frequencies that indicate how many times each subsequence appears in the input.
\\ Different scalar functions can now be used as measure of information. As long as it is a matrix nondecreasing function \cite{Boyd2004} the problem remains convex and the global optimum can be computed. In this paper we consider  D-optimal designs, meaning that the determinant of the Fisher Information matrix is maximized.  In addition we  have opted for a dispersion-based method, which was already successfully applied in the linear case \cite{Schoukens&Pintelon1991}. Advantages of this method are its intuitive interpretation, straightforward implementation and monotonic convergence to the global minimum. In addition we will illustrate with  a short numerical example that the dispersion algorithm can compete with general purpose convex optimizers like cvx \cite{cvx} in terms of computation time.
\\ The minimization of the maximum of the dispersion function yields an  \emph{optimal relative frequency vector}. As already explained, this optimal frequency vector  may not correspond to a realizable time sequence. To alleviate this problem, constraints need to be incorporated into the optimization problem. The space of frequency vectors which satisfies these constraints forms a polyhedron \cite{Patricio_Automatica}. This allows one to represent every frequency vector in the search space as a convex combination of a set of corner points. Unfortunately, computing this set of corner points is numerically expensive. Therefore, an alternative way that approximates this set is proposed in this paper.  It is based on restricting the search space to a constrained set of non-overlapping symmetric frequency vectors. 
\\ Once a realizable and optimal frequency vector is found, a time sequence satisfying this design needs to be derived. To this end a graph-based method  was  suggested in \cite{Larsson2010} and later elaborated in \cite{Patricio_Automatica} for stochastic input sequences. In this work a similar graph-based method is presented for deterministic input sequences.
The problems addressed in this paper have close connections with those addressed in \cite{Larsson2010,Patricio_Automatica} and \cite{Forgione2014}. The main difference with  \cite{Larsson2010,Patricio_Automatica} lies in the fact that in these papers  stochastic inputs are considered and that  the problem is parametrized with respect to the probability density of the subsequences, instead of their relative frequency. While this makes little difference when it comes to the numerical computation of the designs, the interpretation of the results is quite different. 
Moreover, in \cite{Patricio_Automatica} no relation was made between 
the memory of the system and the length of the subsequences used in the input generation method.  We study this relation in  Section \ref{app}, where we show that the designed input is only optimal if both memories are equal. If the memory of the generation method is shorter than the memory of the system, the search space becomes too restrictive. If the memory of the generation is chosen longer, the same results are obtained but at a higher computational cost.
\\ In \cite{Forgione2014} an optimal deterministic input is computed for the class of fading memory nonlinear systems. However, unlike this paper, no sequence generation method was presented in \cite{Forgione2014}.
\\ In addition to  the use of a simple dispersion-based criterion,  the use of an alternative search space spanned by a non-overlapping symmetric set that considerably reduces the computation time,  another novel contribution of this paper is the discussion of the interaction between the system memory and the subsequence length.
\emph{Overview}\\
The paper is structured as follows. Section \ref{sec:Problem} formalizes the optimal input design problem for the considered class of systems. Section \ref{sec:Solution} shows how the associated optimization problem can be solved based on the dispersion function. In Section \ref{sec:Signal} the problem of signal generation is considered, and a graph-based method is proposed. Section \ref{sec:Constraint} discusses how the constraints needed for signal generation can be incorporated into the optimization. Section \ref{sec:Example} illustrates our method on a numerical example. In Section \ref{sec:complexity} the computation time of the dispersion-based method is compared with cvx. In Section \ref{app} we motivate why the subsequence length should be chosen equal to the memory length of the system. Section \ref{sec:Conclusion}  summarizes the obtained results.
\section{Problem Statement} \label{sec:Problem}
The goal of this paper is to find the most informative input of given length $N$ for a nonlinear FIR-type system (as defined in Assumption 1 below) with a known model structure, and disturbed with independent Gaussian output noise. As a measure of information the determinant of the Fisher information matrix is considered. Therefore the design is called D-optimal. When the parameters are identified with such an input sequence, and the estimator is assumed efficient, the volume of the uncertainty ellipse in the parameter space is minimal \cite{Federov1972}. The following assumptions describe this problem formally.
\begin{assumption} \label{ass:systemclass}
The considered system is a member of the class of nonlinear FIR-type systems with memory length $n$ and which are differentiable with respect to the parameters of the system. This model class was first studied in \cite{Larsson2010} in the context of optimal input design.
\begin{equation} \label{modelclass}
 y_0(t,\ta)=G_{NL}(u_0(t),u_0(t-1),..,u_0(t-n+1),\theta)
\end{equation}
where $u_0(t)$ is the noiseless input, $y_0(t,\ta)$ is the noiseless output, $\theta \in \mathbb{R}^{N_{\theta}}$ are the parameters of the model. Notice that the output at time $t$ only depends on the current input sample and $n-1$ previous input samples. 
\\Additionally it is assumed that the system is identifiable with respect to the parameters, meaning that there exists an input sequence $u(1),\ldots, u(N)$ such that the outputs $\{y_0(t,\ta), t=1,\ldots,N\}$ and $\{y_0(t,\ta_1), t=1,\ldots,N\}$ of the corresponding models (\ref{modelclass}) are identical only if ~$\ta_1=\ta$.
\end{assumption}
\begin{assumption}
The class of inputs will be restricted to deterministic time sequences with a length of $N$ samples, whose amplitude can only take values from a finite, predefined set of $A$ values:
\vspace{-2mm}
\begin{eqnarray}
\forall t:~ u(t)\in \{u_1,...,u_A\}
\end{eqnarray}
\end{assumption}
\begin{assumption}
\label{ass:noise}
The output $y(t)$ of the true system is obtained as the sum of a noise-free output $y_0(t,\ta_0)$ defined by (\ref{modelclass}) with a ``true'' parameter vector $\ta_0$ and some 
 additive independent identically distributed (i.i.d) Gaussian noise $e(t)$. The noise is also independent of the input signal $u_0$.
 \vspace{-2mm}
\begin{eqnarray}
\nonumber u(t)&=&u_0(t) \\
\nonumber y(t)&=&y_0(t,\ta_0)+e(t)\\
\nonumber e(t) &\sim& N(0,\sigma^2)
\nonumber
\end{eqnarray}
\end{assumption}
{\bf Criterion} \label{ass:criterion}
The D-optimality criterion is used, which means that the optimal input sequence $u_{opt}$ of length $N$ corresponds to the sequence for which the determinant of the  information matrix $\Fi$ is maximal: 
\vspace{-2mm}
\begin{eqnarray*}
u_{opt}&=&\arg\limits_{u}\max(\mathrm{det}(\Fi))
\end{eqnarray*}
\begin{remark}
Since  the noise variance $\sigma^2$ only scales the Fisher information matrix, it will not alter the optimal input design.
\end{remark}
Given the noise assumption, $\Fi$ can be computed based on the time domain data as (see \cite{SIFDA}):
\begin{eqnarray} \label{eq:Fisher}
\Fi&=& \frac{1}{\sigma^2} \left( \frac{\partial y_0}{ \partial \theta} \right)^T \left( \frac{\partial y_0}{ \partial \theta} \right)
\end{eqnarray}
where $\frac{\partial y_0}{\partial\theta}$ is a $N \times N_{\theta}$ matrix containing the partial derivatives of $y_0$, and $V^T$ stands for the transpose of $V$. 
\begin{remark}
For the computation of the optimal input, it will be assumed that the true parameters of the system,  $\ta_0$, are known. While this may seem in contradiction with the final goal of system identification it is a standard assumption in the field of optimal input design \cite{Federov1972}.
\end{remark}
\section{Problem Solution} \label{sec:Solution}
In order to solve the D-optimal input design problem, as presented in the previous section, three important steps will be made. First, the concept of the $n$-length subsequences is formally introduced. Second, it will be shown that the Fisher information matrix can be expressed as a weighted sum of the Fisher matrices associated to each subsequence. This property is the key to solve the optimization problem efficiently. Third, it will be shown how the problem can be solved with a dispersion-based method similar to the one used in the linear case, as described in \cite{Goodwin&Payne1977,Schoukens&Pintelon1991}.
\subsection{Subsequences}
Considering the system model, it is clear that each output sample of the system depends only on a subsequence $(u(t),u(t-1),...u(t-n+1))$ of the input sequence. 
\begin{definition}
A subsequence is an ordered set of $n$ values each, of which is drawn out of the predefined set $\left\{u_1,u_2,...u_A\right\}$. In total $A^n$ different subsequences can be defined. The space that contains all possible subsequences will be called $C \in \left\{ \mathbb{R}^n \right\}^{A^n}$.
\end{definition}
\begin{notation} \label{notation1}
Each index set $(i_1,...,i_n)$ with $i_1,i_2,...i_n \in \left\{1,2,..,A\right\}$ can be made to correspond to a unique integer  index $k$ defined as follows:
\vspace{-2mm}
\begin{eqnarray} 
k \bydef i_{1}+\sum_{k=2}^{n}(i_{k}-1) \cdot A^{(k-1)}   \label{eq:c}  
\end{eqnarray}
Notice that $(i_n,i_{n-1}, ...,i_1)$ is the representation of $k$ in base $A$,  that $k$ ranges from $1$ to $A^n$, and that (\ref{eq:c}) defines a one-to-one mapping between $k$ and the index set $(i_1,i_2,...,i_n)$.  Thus, the mapping (\ref{eq:c}) establishes the equivalence:
\be
k \Longleftrightarrow (i_1,i_2,...,i_n) \label{equivalence}
\ee
In the remainder of the paper we use indistinctly the notation 
$f(i_1,i_2,...,i_n)$ or $f(k)$ for a function $f(.)$ of the index set,  where $k$ and $(i_1,i_2,...,i_n)$ are related by (\ref{eq:c}).
\end{notation}
\begin{notation} \label{notation2}
In order to be able to refer to a specific subsequence from $C$, the following notation is introduced:
\begin{eqnarray}\label{Cnsubsequence}
c(i_1,i_2,...,i_n)&=&(u_{i_{1}},u_{i_{2}},...,u_{i_{n}})  \nonumber \\
\forall i_1,i_2,...i_n &\in& \left\{1,2,..,A\right\} \nonumber
\end{eqnarray}
Applying the mapping introduced in Notation \ref{notation1}, leads to the following shorthand notation $ c(k) \bydef c(i_1,i_2,...,i_n)$.
\end{notation}

\subsection{Fisher information and frequency vector}
\begin{figure} [thpb] 
	\centering{
		\includegraphics[scale=1.0]{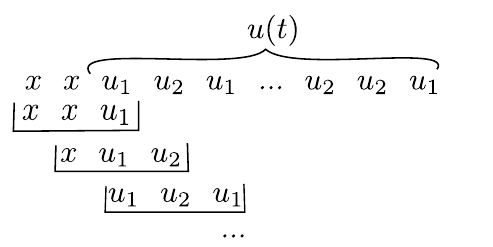}
		}
	\caption{Example of how the subsequences are counted in the case of a FIR-type system with $n=3$ and two amplitude values $\{u_1,u_2\}$. x represents an unknown sample which should be resolved by the chosen initial conditions.}
\label{fig:tupcount}
\end{figure}

From (\ref{eq:Fisher}) it is clear that the $ij^{th}$ element of the Fisher information matrix can be written as a sum over the time samples:
\begin{eqnarray} \label{eq:timesum}
~\Fi(i,j)=\frac{1}{\sigma^2}\sum_{t=1}^{N}{f_{i,j}(t,u,\theta)}
\end{eqnarray}
where the function $f_{i,j}(t,u,\theta)$ corresponds to the product of the partial derivatives of $y_0$ with respect to the $i^{th}$ and $j^{th}$ parameter, evaluated at time instant $t$ for a given input sequence $u$.
\\Because the model class was restricted to FIR-type systems the functions $f_{i,j}$ depend at most on $n$ successive input values.
\begin{eqnarray} 
f_{i,j}(t,u,\theta)=f_{i,j}(u(t),u(t-1),...u(t-n+1)),\theta)
\end{eqnarray}
In other words, the function $f_{i,j}(t,u,\theta)$ depends on the subsequence that ended at time t.
\begin{remark} \label{re:periodicity}
Notice that the first $n-1$ terms depend upon samples which are not measured. Their value is set by the initial conditions. It will be assumed that the signal is periodic with period $N$. This allows us to determine the values of these unknown samples.
\end{remark}
Since the number of possible subsequences is fixed, the number of different terms in (\ref{eq:timesum}) is also fixed. If we compute the values of $f_{ij}$ for each possible subsequence we can reorder the sum over time such that we obtain a weighted sum over all possible subsequences:
\begin{eqnarray} \label{eq:Fi_rearranged}
\Fi(i,j)&=&\frac{1}{\sigma^2}\sum_{k=1}^{A^n}{\xi_N(k) \cdot f_{i,j}(c(k),\theta)} \nonumber\\
&=& ~\sum_{k=1}^{A^n}{\xi_N(k)M_k(i,j)}
\label{eq:sumOversubsequences}
\end{eqnarray} 
The weight $\xi_N(k)$ indicates how often the  subsequence $c(k)$ 
occurs in the sequence $u(t)$ and is therefore called the frequency of the sequence $c(k)$.
From (\ref{eq:Fi_rearranged}) it is clear that the Fisher information matrix for a given input sequence $u$ is completely determined by the subsequences that $u$ contains.
\begin{definition}
The Fisher information matrix $M_k$ corresponding to the $k^{th}$ subsequence will be called the $k^{th}$ elementary Fisher matrix:
\begin{eqnarray} \label{definfom}
\Fi_k(i,j)=\frac{1}{\sigma^2}\left( \frac{\partial y_0}{ \partial \theta_i} \right)^T\left( \frac{\partial y_0}{ \partial \theta_j} \right)|_{c(k)}
\end{eqnarray}
where the partial derivatives are evaluated for the subsequence $c(k)$.
\end{definition}
\begin{definition}
The vector $\xi_{N}(\cdot)\in\mathbb{N}^{A^n}$, containing the number of times each subsequence occurs in the input design, is called the frequency vector of the design. 
\end{definition}
\begin{remark}
Given a time sequence $u(t)$, the corresponding frequency vector can be obtained by counting the number of times each subsequence occurs in the signal $u(t)$. The subsequences are counted as depicted in Fig.\ref{fig:tupcount}. Notice that a signal with N samples contains N subsequences, due to the assumed periodicity of the signal (see Remark \ref{re:periodicity}).
\end{remark}
Normalizing  (\ref{eq:sumOversubsequences})  by the number of subsequences allows us to rewrite the normalized Fisher information matrix as a convex combination of the elementary matrices with the relative frequencies as convex coefficients:
\begin{eqnarray} \label{Mdecomp}
\frac{\Fi(\xi_N)}{N}&=&\sum_{k=1}^{A^n}{\xi(k).\Fi_k}  
\end{eqnarray} 
Notice that only the coefficients $\xi(k)$ depend upon the particular design used in the input $u(t), t=1,\ldots, N$. The elementary information matrices $\Fi_k$ are independent of the design and can be computed a priori given the amplitude set $A$ and the memory $n$ of the FIR type nonlinear system.
\begin{definition} 
The frequency vector divided by the total number $N$ of subsequences in the design is called the relative frequency vector $\xi(\cdot)\in\mathbb{R}_+^{A^n}$. By construction the normalized frequencies have the properties of convex coefficients, meaning that their values range from 0 to 1 and that their sum is exactly one:
\begin{eqnarray} \label{eq:convexCoeff}
&&~~~~~\forall k: ~~~~ \xi(k)=\xi_N(k)/N \nonumber 
\\ &&~~~~\xi(k) \in [0,1] ~~~ \mbox{and} ~~~ \sum^{A^n}_{k=1} \xi(k)=1
\end{eqnarray}
\end{definition}
\begin{remark}
By performing the optimization for the relative frequency vector $\xi$, we relax the problem by making the search space continuous. However, the frequency vector $\xi_N$ can only contain natural numbers. So, after denormalizing the frequency vector, the values will be rounded to the nearest natural number. This may cause a slight decrease in information.
\end{remark}
Computing an optimal experiment consists of finding the vector $\xi$ that maximizes the determinant of the normalized information matrix:
\begin{eqnarray*}
\xi_{opt}&=& \arg\limits max_{\xi}(det(\frac{\Fi(\xi)}{N})) 
\end{eqnarray*}
\subsection{Dispersion function}
In the previous subsection it was shown that the normalized Fisher information matrix can be rewritten as a convex combination of known elementary information matrices. We now show that this allows us to use a dispersion-based method in order to perform the optimization. Instead of solving the optimization problem directly, an equivalent problem is solved where the maximum of an auxiliary function, called the {\it dispersion function}, is minimized. The dispersion function, also called {\it response dispersion}, was introduced in the experiment design for identification arena in \cite{Goodwin&Payne1977}.
\begin{definition}
With the notations introduced above, the dispersion function $v(.,.)$ is defined as:
\vspace{-2mm}
\begin{eqnarray}
v(\xi,k)=\mathrm{trace}(\Fi(\xi)^{-1}\cdot \Fi_k)
\label{eq:Dispersion}
\end{eqnarray}
where $\Fi(\xi)$ is the information matrix computed for the given design $\xi$, and $\Fi_k$ is the information matrix corresponding to the $k^{th}$ subsequence.
\end{definition}
Some useful properties of the dispersion function are \cite{Goodwin&Payne1977}:
\vspace{-2mm}
\begin{itemize}
	\item The maximal value of the dispersion can never be smaller than the number of independent parameters in the model $N_\theta$. 
	\item For any design $\xi$, the inner product  $\sum_{k=1}^{A^n}{v(\xi,k) \cdot \xi(k)}$  equals the number of free parameters in the model. 
	\item The dispersion function can also be interpreted as a normalized variance of the estimated model.
\end{itemize}
\begin{theorem}
\label{theo:dispersion}
The following characterizations of an optimal design are equivalent:
\vspace{-2mm}
\begin{enumerate}
	\item $\xi_{opt}~$ maximizes $~\mathrm{det}(\Fi)$
	\item $\xi_{opt}~$ minimizes $~~\mathrm{max_{k}} v(\xi,k)$
	\item $~\mathrm{max_{k} }v(\xi_{opt},k)=N_{\theta}$
\end{enumerate}
where $N_{\theta}$ is the number of independent parameters in the model.\\
{\bf Proof:} see \cite{Goodwin&Payne1977} Chapter 6, page 147.
\end{theorem}
Theorem \ref{theo:dispersion} states that the design that maximizes the determinant of the Fisher information matrix is the same design that minimizes the maximum of the dispersion function. Since a simple and efficient algorithm exists that solves the latter problem, we shall adopt it for the computation of our optimal experiment.
\subsection{Optimization Algorithm} \label{subsec:optimization}
In \cite{Schoukens&Pintelon1991} a simple and stable, monotonically converging algorithm is presented, which finds the design that minimizes the maximum of the dispersion function. This algorithm can be summarized in four steps: 
\vspace{-2mm}
\begin{enumerate}
\item Initialize with a uniform design: $\xi(k)=1/A^n$ 
\item Compute the dispersion function $v(\xi,k)$  for the current design using (\ref{eq:Dispersion})
\item Update the design in accordance with the dispersion function as follows $\xi_{new}(k)=\frac{v(\xi, k)}{A^n}.\xi_{old}(k)$ 
\item Stopping criterion: if $(\mathrm{max_k}v(\xi_{new},k)-N_{\theta})$ is smaller than a predefined threshold, the optimal solution is assumed to be found; else go to step 2.
\end{enumerate}
\vspace{-2mm}
The stopping criterion is based on the third expression of Theorem 1. The monotonic convergence of the algorithm is proven in \cite{Yaming2010}. In Section~\ref{sec:complexity}  the performance of this dispersion-based algorithm will be compared to the general purpose convex optimizer cvx.
\begin{remark}
Notice that once a particular frequency $\xi(k)$ becomes zero for some $k$, this frequency remains zero for all subsequent iterations. Therefore, the computational speed of the algorithm can be improved by only updating the nonzero frequencies. This avoids unneeded evaluations of the dispersion function.
\end{remark}
\section{Signal Generation} \label{sec:Signal}
The optimal frequency vector $\xi_{N,opt}$  can be interpreted as an experiment of $N$ measurements, where each measurement consists of applying a single subsequence to the system and measuring the corresponding output sample.
\\A naive way to perform the optimal design is to concatenate all the subsequences contained in $\xi_{N,opt}$ (i.e. all the subsequences $c(k)$ for which $\xi_{N,opt}(k) \neq 0$), and only measure the output samples which correspond to these subsequences. This means that a signal with $nN$ samples is used at the input, in order to collect $N$ samples at the output. Clearly this is not an efficient approach, since only $N$ out of the $nN$  output samples are used for parameter estimation.
\\It would be better to generate a periodic input sequence with a period length of $N$ samples, containing the $N$ needed subsequences. However, not every frequency vector has a corresponding input sequence of length $N$ because the $n-1$ last inputs of subsequence $c(k)$ for which $\xi_{opt}(k)\neq 0$ may not correspond to the $n-1$ first inputs of another subsequence $c(j)$ for which $\xi_{opt}(j)\neq 0$.
In order to derive conditions on the frequency vector which guarantee the existence of at least one realizable time sequence, a sequence generation method will be introduced. This generation method will correspond with a path through the associated graph of the frequency vector. 
\begin{definition}
Given a subsequence of length n, the right subsubsequence is defined as the subsequence of length n-1 obtained by removing the first element of the original subsequence. Similarly, the left subsubsequence is defined as the subsequence of length n-1 obtained by removing the last element of the original subsequence.
\end{definition}
\begin{definition}
The graph associated with a frequency vector $\xi_N$ (see Notation \ref{notation1}) is defined as follows:
\begin{itemize}
	\item The graph  contains $A^{n-1}$ nodes, each containing a different subsubsequence of length n-1
	\item Each subsequence of length n corresponds to a directed edge connecting its left and right subsubsequence. The edge  starts in the left subsubsequence and  ends in the right one.  
	\item Each edge has a multiplicity which corresponds to the subsequence frequency $\xi_N(i_1,\ldots,i_n)$ of its corresponding subsequence.
\end{itemize}
\end{definition}
\begin{example}
Consider a FIR-type system with $n=3$ and two amplitude levels $\{u_1,u_2\}$. In total $2^{(3-1)}$ different subsubsequences of length n-1 can be defined. So the associated graph contains four nodes. Additionally $2^{3}$ different subsequences of length n can be defined. This means that the graph  contains eight edges. For example, the edge corresponding to the subsequence $u_1u_1u_2$ connects its left subsubsequence $u_1u_1$ with its right subsubsequence $u_1u_2$. The multiplicity of the edge connecting $u_1u_1$ with $u_1u_2$ equals its frequency $\xi_N(1,1,2)$. If this reasoning is repeated for every subsequence, the graph in Fig.\ref{fig:graph} is obtained.
\begin{figure} [thpb] 
\begin{center}
\includegraphics[scale=1.0]{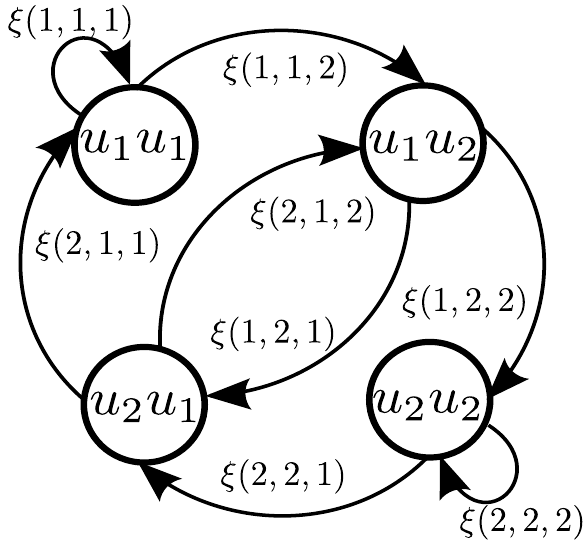}
\caption{Associated graph in the case of $n=3$ and two amplitude values}
\label{fig:graph}
\end{center}
\end{figure}
\end{example}
If there exists a time sequence corresponding to the frequency vector $\xi_N$, then this sequence can be obtained by the following  steps:
\vspace{-2mm}
\begin{enumerate}
	\item Construct the graph associated with the frequency vector $\xi_N$
	\item Find a path,  starting from an arbitrary node, that uses each edge exactly as many times as its multiplicity indicates. 
	\item For every edge in the path, add the last amplitude value of the corresponding subsequence to the end of the input sequence.
\end{enumerate}
From the above, it can be concluded that a time sequence exists if there exists a path, starting from an arbitrary node, that uses each edge exactly as many times as its multiplicity indicates. In graph theory such a path is called an Euler cycle. Some well known algorithms for finding the Euler cycle (if it exists) are the algorithm of Fleury \cite{Fleury1883} and the algorithm of Hierholzer. More recent versions of these algorithms can be found in any textbook on graph theory \cite{GraphRef}. 
\\In order for an Euler cycle to exist, the associated graph can not be disjoint and all vertices need to have a zero degree, which means the sum of multiplicities for all outgoing edges minus the sum of multiplicities for all incoming edges needs to be zero. 
\begin{theorem} 
A periodic time sequence exists that realizes a prescribed frequency vector $\xi_N$ only if its associated graph is not disjoint and the frequency vector satisfies the following conditions: 
\begin{eqnarray}
\label{eq:Constraint}
		\forall i_{1},i_{2},...,i_{n-1} &\in& \{1,...,A\} \nonumber \\
		\sum_{j=1}^{A}\xi_N(j,i_{1},...,i_{n-1})&=&\sum_{j=1}^{A}\xi_N(i_{1},...,i_{n-1},j) 
\end{eqnarray}
or equivalently, using the scalar index $k$ rather than the vector index  $(i_1, \ldots, i_n)$, as defined in Notation 1:
\begin{eqnarray} 
\forall m \in [1, \ldots, A^{n-1}]:&& \nonumber\\
 \sum_{j=1}^{A}\xi_N(j+(m-1)A)&=&\sum_{j=1}^{A}\xi_N(m+(j-1)A^{n-1}) \label{equivconstraint}
\end{eqnarray}
\noindent {\bf Proof:} A time sequence  has the correct frequency vector $\xi_N$ if a path can be constructed from the associated graph whereby each edge is used exactly as many times as its multiplicity indicates.
In order for such a path to exist, the sum of the multiplicities of the outgoing and incoming edges need to be equal in every node, where each node is  defined by a $(n-1)$ vector index $(i_1, \ldots, i_{n-1})$. This is precisely the constraint (\ref{eq:Constraint}). 
Now fix a $(n-1)$ subsequence $(i_1, \ldots, i_{n-1})$, and let $m$ denote the scalar index for this $(n-1)$ subsequence, i.e.
\beqnd
m = i_1 + (i_2-1)A + (i_3-1)A^2 + \ldots + (i_{n-1}-1)A^{n-2}
\eeqnd
We now express the two indices of length n appearing in (\ref{eq:Constraint}) 
as a function of $m$:
\beqnd
(j,i_1, \ldots, i_{n-1}) &=& j + (i_1 - 1)A + (i_2-1)A^2\\
&&+ \ldots + (i_{n-1}-1)A^{n-1} \\
&=& mA- A + j= j+ (m-1)A 
\eeqnd
With exactly the same procedure one gets
\beqnd
(i_1, \ldots, i_{n-1}, j) = m +(j-1)A^{n-1}
\eeqnd
\end{theorem}\begin{remark}
In order to illustrate that the equations in (\ref{eq:Constraint}) are not sufficient conditions for the existence of a realizable time sequence for a given a frequency vector,  consider a disconnected graph which satisfies the constraints. In such a graph there is no single path connecting all the nodes, meaning there is also no corresponding time sequence.
\end{remark}
\begin{remark}
In a stochastic framework, the frequency matrix can be considered as mutual discrete probability distribution functions of the n stochastic variables. Imposing that the signal is stationary, will lead to the same constraint as given in (\ref{eq:Constraint}) \cite{Larsson2010}. The graph described above can then be seen as a Markov chain used to generate a realization of the frequency matrix.
\end{remark} 
\section{Constrained Optimization} \label{sec:Constraint}
In the previous section it was shown that a frequency vector can only be realized as a time sequence if it meets the conditions (\ref{eq:Constraint}). Therefore, these conditions will be imposed during the optimization. Unfortunately the dispersion-based algorithm cannot handle constraints directly. Instead, the realizable search space will be represented as a convex set and the optimization will be performed with respect to the new convex coefficients.
\\We now show that the set of relative frequency vectors meeting the condition (\ref{eq:Constraint}) are a convex set. First, it should be noted that the relative frequencies are convex coefficients (see (\ref{eq:convexCoeff})). Therefore the full search space is contained in a convex polyhedron \cite{Boyd2004}. Second, the constraints presented in (\ref{eq:Constraint}) correspond to a set of linear equality constraints, meaning they describe a subspace centered at the origin. Combining these two observations shows that the space of realizable relative frequency vectors consists of the intersection between a polyhedron and a subspace, which in turn is again a polyhedron \cite{Boyd2004} and therefore a convex set.
\\Knowing that the search space is a convex polyhedron allows us to express every realizable relative frequency vector as a convex combination of the corner points:
\begin{eqnarray}\label{gammadecomp}
&&\forall k \in \{1,2,\ldots, A^n\}: ~\xi_{\gamma}(k)=\sum_{j=1}^{N_{b}}\gamma(j).\xi_{b_{j}}(k) \\
&& \mathrm{with} ~ \gamma(j)\in[0,1] \mathrm{~and~} \sum_{j=1}^{N_b}\gamma(j)=1 \nonumber
\end{eqnarray}
where $\xi_{b_{j}}(k)$ are the mentioned corner points, $N_b$ is the number of corner points, $\gamma(j)$ are the new convex coefficients with respect to which the optimization will be performed, and $\xi_{\gamma}$ is the frequency vector corresponding to the coefficient vector $\gamma$.
\\In \cite{Patricio_Automatica} the same reasoning was followed. Additionally, it was shown how the corner points that span the polyhedron could be constructed from the elementary cycles present in the graph associated to the unity frequency vector. However finding these elementary cycles is a computationally heavy task and becomes infeasible for medium sized graphs \cite{Hawick2008}.
\subsection{Non-overlapping Symmetric Set}
As an alternative, we present a more restrictive convex set that has the advantages that its corner points can easily be computed. However, it can not be guaranteed that this set contains the global optimum of the polyhedron of realizable frequency vectors.
\begin{definition} \label{def:nonoverlappingsymmetric}
A  set of vectors $[\xi_{b_1},\xi_{b_2},...,\xi_{b_{N_b}}]$ in $\mathbb{R}^{A^n}$is called non-overlapping symmetric if they have the following four properties:
\begin{enumerate}
	\item positivity constraint: \\ $\forall k,~\forall j: \xi_{b_{j}}(k)\geq0$
	\item non-overlapping constraint: \\ $\forall k,\,\exists!j:\, \xi_{b_{j}}(k)>0$
	\item unity sum constraint: \\$\forall j:\, \sum_{k=1}^{A^n}{\xi_{b_{j}}(k)}=1$
	\item symmetry constraint: 
\\$\forall j, ~ \forall i_1,...,i_n, ~\forall p\in\ \mathrm{Perm}_{1,2,...,n}:$
\\$\xi_{b_{j}}(i_{1},i_{2},...,i_{n})=\xi_{b_{j}}(i_{p_1},i_{p_2},...,i_{p_n})$
\end{enumerate}
where $N_b$ indicates the number of vectors in the set, $\mathrm{Perm}_{1,2,...,n}$ stands for the set of all possible permutations of the symbols $\{1,2,...,n\}$, and $\exists!j$ means 'there exists only one'.
\end{definition}
\begin{remark}
The first property guarantees that the frequencies are nonnegative. The second property guarantees that the vectors don't have the same nonzero element positions, thereby making these vectors linearly independent.  The third property ensures that the sum of the frequencies is one. The first three properties are needed in order to impose (\ref{eq:convexCoeff}) on $\xi_\gamma$. The fourth property imposes symmetry on the frequency vector. This symmetry constraint implies that the constraint (\ref{eq:Constraint}) is satisfied.
\end{remark}
\begin{remark}
The number of non-overlapping symmetric vectors $N_b$ equals the number of degrees of freedom in a symmetric tensor of order $n$ and dimensionality $A$. For example if $n=2$, $N_b= \frac{A(A+1)}{2}$, which correponds to the degrees of freedom in a symmetric $A\mathrm{x}A$ matrix.
\end{remark}
\begin{example}
In the case of $n=2$ and two possible amplitude values (i.e. $A=2$), the number of vectors in the non-overlapping symmetric set  is $N_b=3$. The non-overlapping symmetric set is given by the three frequency vectors below:
\begin{eqnarray*}
\xi_{b_{1}}=\left[\begin{array}{c}
1 \\ 0 \\ 0\\ 0
\end{array}\right],\, \xi_{b_{2}}=\left[\begin{array}{c}
0 \\ 0.5\\ 0.5 \\ 0
\end{array}\right],\, \xi_{b_{3}}=\left[\begin{array}{c}
0 \\ 0\\ 0 \\ 1
\end{array}\right] 
\end{eqnarray*} 
It is easy to check that the four constraints are satisfied. Note that the fourth constraint reduces here to $\xi_{b_j}(1,2) = \xi_{b_j}(2,1)$ for $j=1,2,3$ or, equivalently $\xi_{b_j}(2) = \xi_{b_j}(3)$ when the unique index $k$ of (\ref{eq:c}) is used in lieu of $(i_1, i_2)$.
\end{example}
\subsection{Symmetric Designs}
Assuming the use of a non-overlapping symmetric set, the expression  of the normalized information matrix can be rewritten as a function of the weighting coefficients $\gamma(j)$ by substituting (\ref{gammadecomp}) into (\ref{Mdecomp}):
\begin{eqnarray} 
&&\frac{\Fi}{N}=\sum_{j=1}^{N_b}{\gamma(j) \sum_{k=1}^{A^n}{\xi_{b_j}(k) \cdot \Fi_k}} =\sum_{j=1}^{N_b}{\gamma(j) \cdot \Fi_{\gamma(j)}}
\label{eq:changeDesigns}
\end{eqnarray}
where $\Fi_{\gamma(j)}=\sum_{k=1}^{A^n}{\xi_{b_j}(k) \cdot \Fi_k}$ are the new elementary information matrices.
\\During the constrained optimization, the dispersion function  will be computed based on these $\Fi_{\gamma(j)}$. We define:
\begin{eqnarray} \label{eq:NewDispersion}
v_{\gamma}(\xi_{\gamma},j)\bydef \mathrm{trace}(\Fi(\xi_{\gamma}(k))^{-1} \cdot \Fi_{\gamma(j)})
\end{eqnarray}
where the subscript ${\gamma}$ indicates that $v_{\gamma}$ is computed from the elementary information matrices $\Fi_{\gamma(j)}$. Notice that the values and dimensions of the dispersion functions  $v_{\gamma}(\xi_{\gamma},j)$ and $v(\xi_{\gamma},k)$ are different. This is because the dispersion function evaluates the quality of the input design relative to the considered search space.
\subsection{Optimization Algorithm Revisited}
By introducing the new elementary information matrices $\Fi_{\gamma(j)}$ and the dispersion function $v_{\gamma}$, and as a result of   the convex properties of the coefficients $\gamma(j)$, the dispersion-based algorithm described in Subsection \ref{subsec:optimization} can be reused to find the $\gamma_{opt}(j)$ for which the determinant is maximal. Revisiting the 4-step algorithm leads to:
\begin{enumerate}
\item Initialize with a uniform design: $\gamma(j)=1/N_{b}$ 
\item Compute the dispersion function $v_\gamma(\xi_{\gamma},j)$ for the current design using (\ref{eq:NewDispersion})
\item Update the design in accordance with the dispersion function as follows $\gamma_{new}(j)=\frac{{v_{\gamma}}(\xi_{\gamma}, j)}{N_b}.\gamma_{old}(j)$ 
\item Stopping criterion: if $(\mathrm{max_j}v_{\gamma}(\xi_{\gamma_{new}},j)-N_{\theta})$ is smaller than a predefined threshold, the optimal solution is assumed to be found; else go to step 2.
\end{enumerate}
\section{Numerical Example} \label{sec:Example}
The methods above will now be illustrated on the following numerical example which consists of a finite impulse response filter, followed by a polynomial nonlinearity:
\begin{eqnarray*}
y(t)&=&c_1w(t)^{3}+c_2w(t)+e(t)
\\w(t)&=&b_1 u(t)+b_2 u(t-1)
\\ u(t)&\in&[-1:2/9:1]~~~\mbox{and}~~~ e(t)\sim N(0,1)
\end{eqnarray*}
with $b= (3; 1)$ and $c=(1; -0.25)$. The value of  $c_2$ will be fixed in order to make the problem identifiable. The noise is zero mean Gaussian distributed with unity variance.
The amplitude set contains 10 uniformly spaced values between -1 and 1 (A=10). Because the system has a memory length of two (n=2), 100 different subsequences should be considered.\\
Notice that this numerical problem leads to a complete graph with 10 nodes and 100 edges. For such graph it is proven that the number of elementary cycles is 1112073 and that computing all these cycles has a complexity of 122328140 \cite{Johnson}. An implementation of Johnson's algorithm \cite{Johnson} in Matlab takes multiple days in order to compute all these elementary cycles. Additionally the optimization problem would need to consider 1112073 variables which is not feasible. In contrast the symmetric design space contains only 55 frequency vectors that can be compute in a couple of seconds.
\subsection{Unconstrained optimization}
\begin{figure} 
	\centering{
	\resizebox{1.0\hsize}{!}{
		\includegraphics[scale=1]{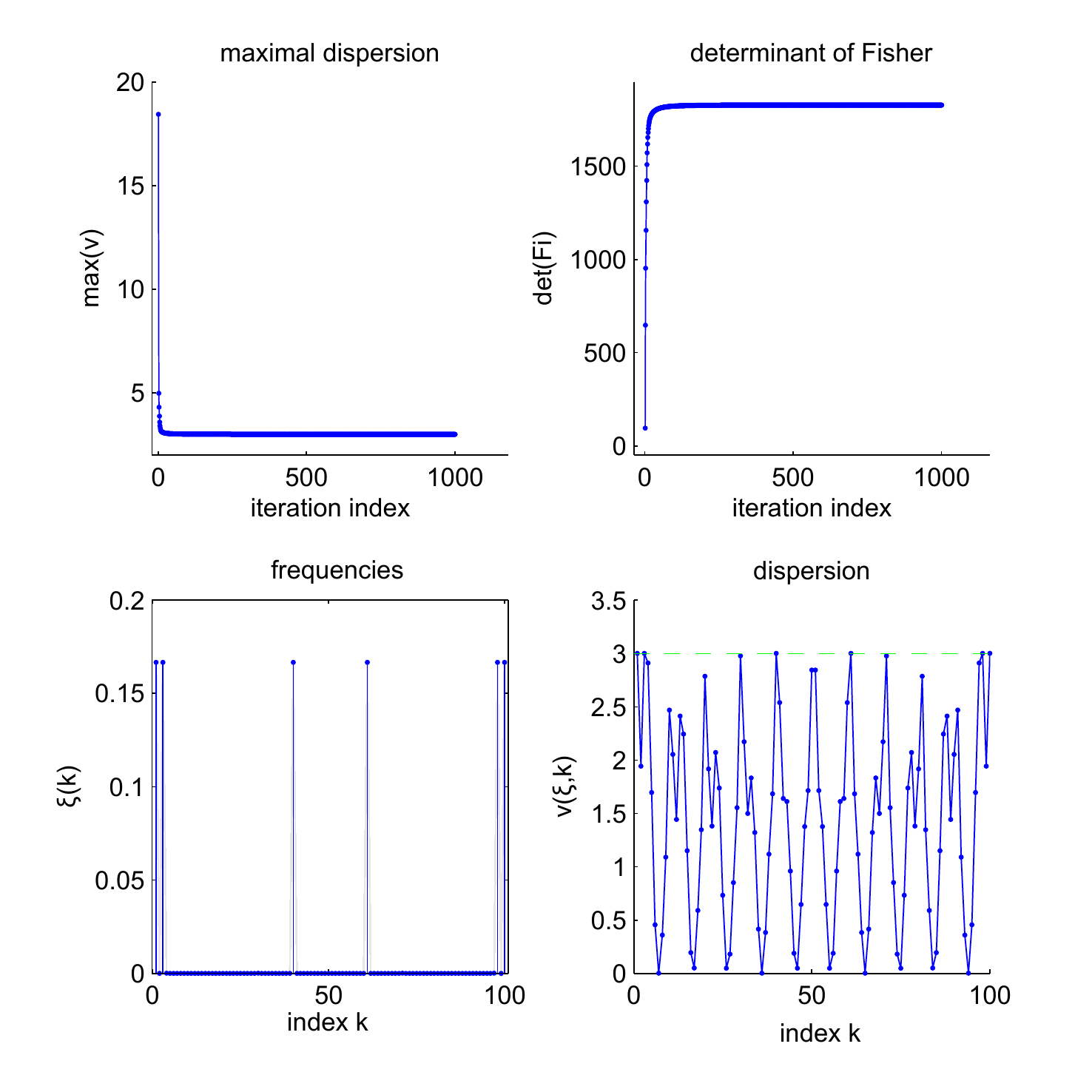}
		}}
	\caption{Results without constraints. Top, left:  maximal dispersion as function of the iteration index Top, right: determinant of the normalized information matrix function of the iteration index.  Bottom,left: relative frequencies. Bottom,right: dispersion function of the optimal design.}
	\label{fig:unconstrained}
\end{figure}

\begin{figure} 
	\centering{
	\resizebox{1.0\hsize}{!}{
		\includegraphics[scale=1]{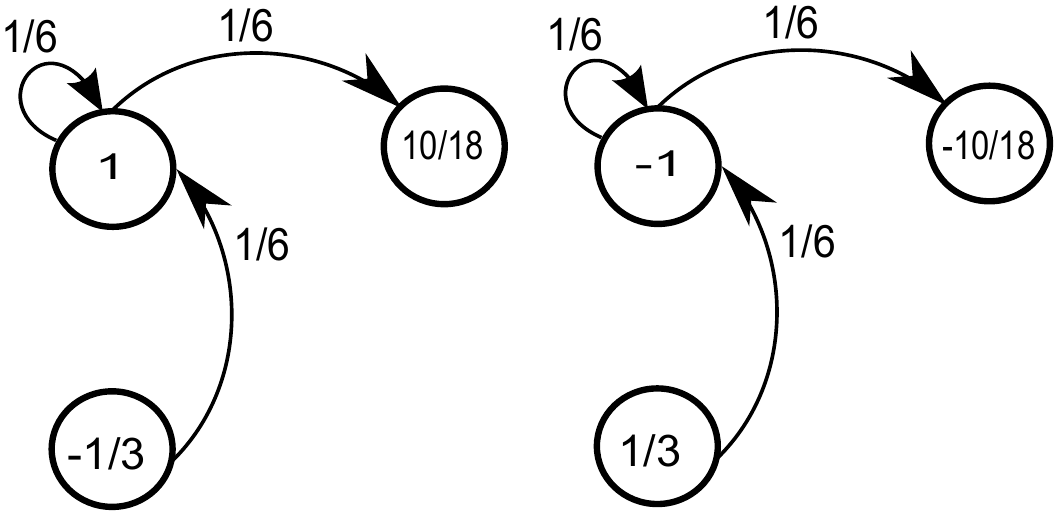}
		}}
	\caption{Associated graph of the optimal unconstrained design.}
	\label{Fig:uncongraph}
\end{figure}

First, the unconstrained optimization is performed. This means that the relative frequencies $\xi(k)$ are optimized with the dispersion-based optimization method described in subsection \ref{subsec:optimization}. The results for 1000 iterations are plotted in Fig.\ref{fig:unconstrained}. In each iteration step the dispersion function, relative frequency vector and determinant of the normalized information matrix are computed and stored. 
\\The top plots represent the evolution of the maximum dispersion and of the determinant of the Fisher information matrix as a function of the iterations. They show that the method successfully lowers the maximal value of the dispersion and at the same time increases the determinant of the information matrix. Note that in the end the maximal dispersion reaches the value of 3, which corresponds to the number of free parameters. This indicates that the obtained solution is optimal (see Theorem \ref{theo:dispersion}). 
\\ The optimal relative frequency vector {at the end of the iterations} is depicted in the bottom left plot; it only contains 6 entries that are different from zero. This means that the optimal design is such that only 6 out of the 100 possible subsequences are used to excite the system. Each of them has the same frequency value. The bottom right plot represents the value of the dispersion function for each of the 100 subsequences.
\\ Table 1 shows the subsequences of the optimal unconstrained design and their corresponding relative frequencies, while Fig. \ref{Fig:uncongraph} represents the associated graph.
Two observations should be made. First, the design does not obey the constraints (\ref{eq:Constraint}). As a result, the frequency vector cannot be realized as a time sequence. Second, the graph is disconnected, which means that not every node can be reached from any other node.
\subsection{Evolution of the Determinant}
Now let us have a closer look at the  evolution of the determinant in Fig.\ref{fig:unconstrained}. For the first 10 to 20 iterations there is a rapid increase in the determinant value. During these iterations, the number of different subsequences is reduced drastically. After this rapid change, the evolution of the determinant value becomes stable. Only the frequencies of the remaining subsequences are changed but the selection of subsequences stays the same. 
From this observation it can be concluded that selecting the optimal subset of subsequences is more important for the quality of the design than finding the optimal frequencies of the selected subsequences.
\begin{table} \label{tab:unconstraint}
\begin{center}
\begin{tabular}{|c|c|c|}
\hline 
index & subsequence & frequency\tabularnewline
\hline 
\hline 
1 & {[}-1;-1{]} & 1/6\tabularnewline
\hline 
3 & {[}-1;-10/18{]} & 1/6\tabularnewline
\hline 
40 & {[}-1/3,1{]} & 1/6\tabularnewline
\hline 
61 & {[}1/3;-1{]} & 1/6\tabularnewline
\hline 
98 & {[}1;10/18{]} & 1/6\tabularnewline
\hline 
100 & {[}1;1{]} & 1/6\tabularnewline
\hline 
\end{tabular}
\end{center}
\label{tab:Unconstraint}
\caption{Table containing the index, subsequences and relative frequencies of the optimal unconstrained design $\xi_{opt}$.}
\end{table}
\subsection{Constrained optimization}
\begin{figure} 
	\centering{
	\resizebox{1.0\hsize}{!}{
		\includegraphics[scale=1]{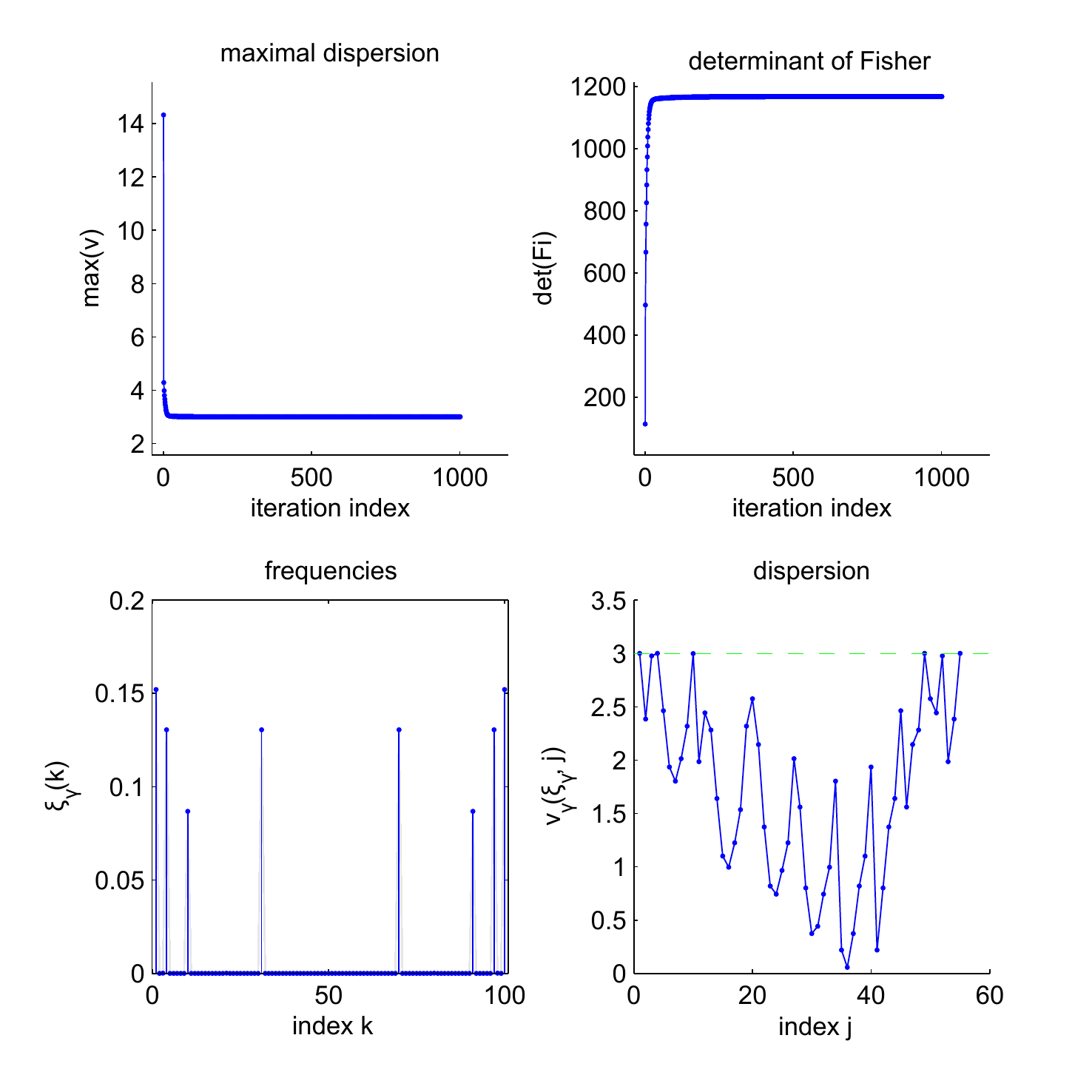}
		}}
	\caption{Results with constraints. Top, left:  maximal dispersion. Top, right: determinant of the normalized information matrix.  Bottom,left: relative frequencies of the optimal design. Bottom,right: dispersion function at the end of the iterations for each of the 55 subsequences.}
	\label{fig:constrained}
\end{figure}

\begin{figure} 
	\centering{
	\resizebox{1.0\hsize}{!}{
		\includegraphics[scale=1]{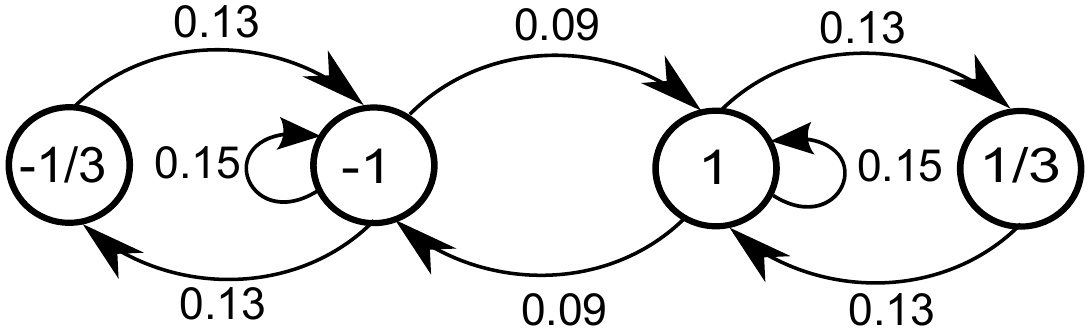}
		}}
	\caption{Associated graph of the optimal constrained design.}
	\label{Fig:congraph}
\end{figure}

Next, the optimization is performed using the non-overlapping symmetric basis vectors obeying the constraints of  Definition~\ref{def:nonoverlappingsymmetric}. 
This means that the optimization is performed with respect to the coefficients $\gamma(j)$ of (\ref{eq:changeDesigns}) using the elementary information matrices $M_{\gamma(j)}$.
 Again 1000 iterations are taken which leads to the plots in Fig.\ref{fig:constrained}.  Due to the symmetry constraint,  only $N_b=55$ coefficients need to be considered; hence $\gamma(j)$ in the bottom right plot ranges from 1 to 55. The determinant increases monotonically while the maximal value of the dispersion $\max(v_\gamma(\gamma_{opt}, j))$ is driven to its minimal value of 3. This indicates that the final design is optimal in its subspace of constrained designs.
\\Table 2 shows the relative frequencies of the optimal constrained design which contains eight different subsequences with different frequencies. As expected, the design is symmetric, meaning that the subsequences $[u_1,u_2]$ and $[u_2,u_1]$ have the same frequency. This allows us to concatenate the subsequences without the need of unwanted transition subsequences, leading to a realizable design that is optimal in the constrained space. The same conclusion can be made from the associated graph in Fig. \ref{Fig:congraph}.
\\
\begin{table}
\begin{center}
\begin{tabular}{|c|c|c|}
\hline 
index & subsequence & frequency\tabularnewline
\hline 
\hline 
1 & {[}-1;-1{]} & 0.15\tabularnewline
\hline 
4 & {[}-1;-1/3{]} & 0.13\tabularnewline
\hline 
10 & {[}-1,1{]} & 0.09\tabularnewline
\hline 
31 & {[}-1/3;-1{]} & 0.13\tabularnewline
\hline 
70 & {[}1/3;1{]} & 0.13\tabularnewline
\hline 
91 & {[}1;-1{]} & 0.09\tabularnewline
\hline 
97 & {[}1;1/3{]} & 0.13\tabularnewline
\hline 
100 & {[}1;1{]} & 0.15\tabularnewline
\hline 
\end{tabular}
\end{center}
\label{tab:Constraint}
\caption{Table containing the index, subsequences and realtive frequencies of the optimal constrained design $\xi_{\gamma_{opt}}$.}
\end{table}
\subsection{Computing the Dispersion}
\begin{figure} 
	\centering{ \resizebox{0.5\hsize}{!}{
		\includegraphics[scale=1]{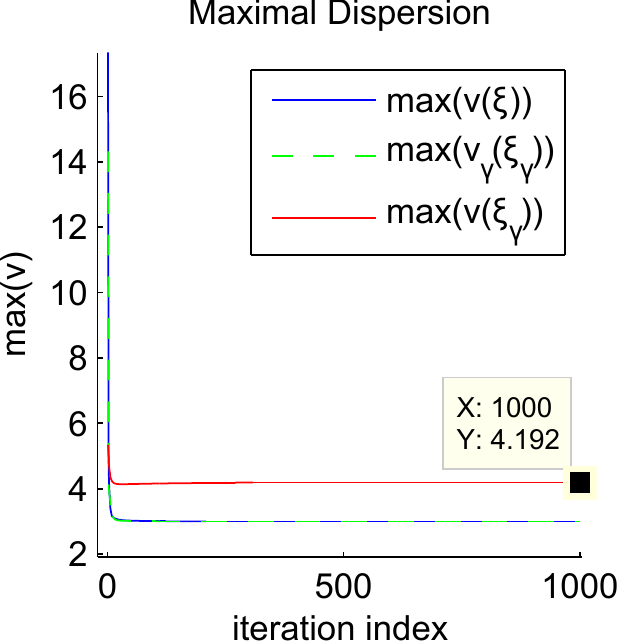}
	}}
	\caption{Maximal dispersion of the normalized information matrix in the constrained case, computed for different elementary designs.}
	\label{fig:corrected}
\end{figure}

At first glance it seems contradictory that the optimal unconstrained and constrained design have the same maximum dispersion but different determinant values. However, the dispersion of the constrained design $v_{\gamma}(\xi_{\gamma},j)$ and the dispersion of the unconstrained design $v(\xi,k)$ can not be compared directly, because they are based on different elementary information matrices. In order to make a comparison possible, the dispersion of the constrained solution $\xi_{\gamma,opt}$ is computed with respect to the elementary Fisher matrices $M_{k}$.
\\The results are plotted in Fig.\ref{fig:corrected}. From the left plot it is clear that the dispersion function $v(\xi_{\gamma})$ is larger than $v_{\gamma}(\xi_{\gamma})$ and $v(\xi)$. This indicates that the constrained solution is not optimal in the full frequency space. This is in accordance with the observation that the determinant of the constrained design $\xi_{\gamma_{opt}}$ is smaller  than the determinant of the unconstrained design $\xi_{opt}$. See Table~\ref{tab:Grid} for the exact determinant values.
\subsection{Signal Generation}
\begin{figure} 
	\centering{
	\resizebox{1.0\hsize}{!}{
		\includegraphics[scale=1]{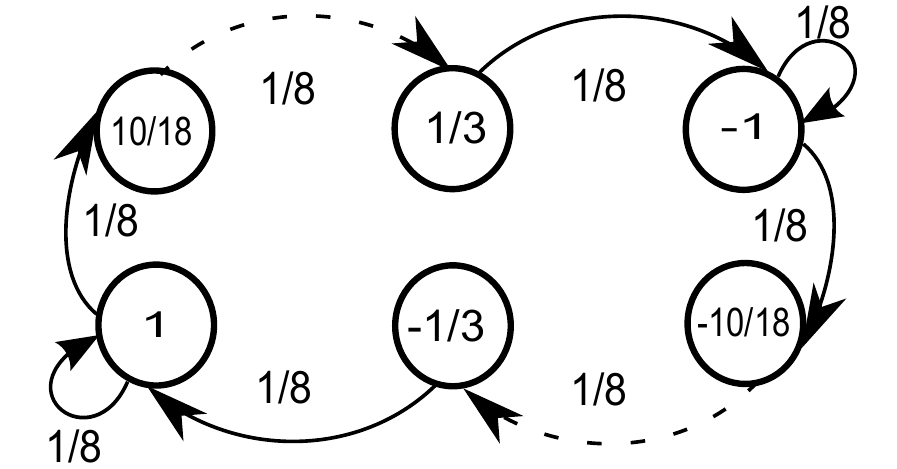}
		}}
	\caption{Modified graph from the unconstrained design. Dotted arrows were added and the relative frequencies were renormalized.}
	\label{Fig:modgraph}
\end{figure}

\begin{figure} 
	\centering{
	\resizebox{1.0\hsize}{!}{
		\includegraphics[scale=1]{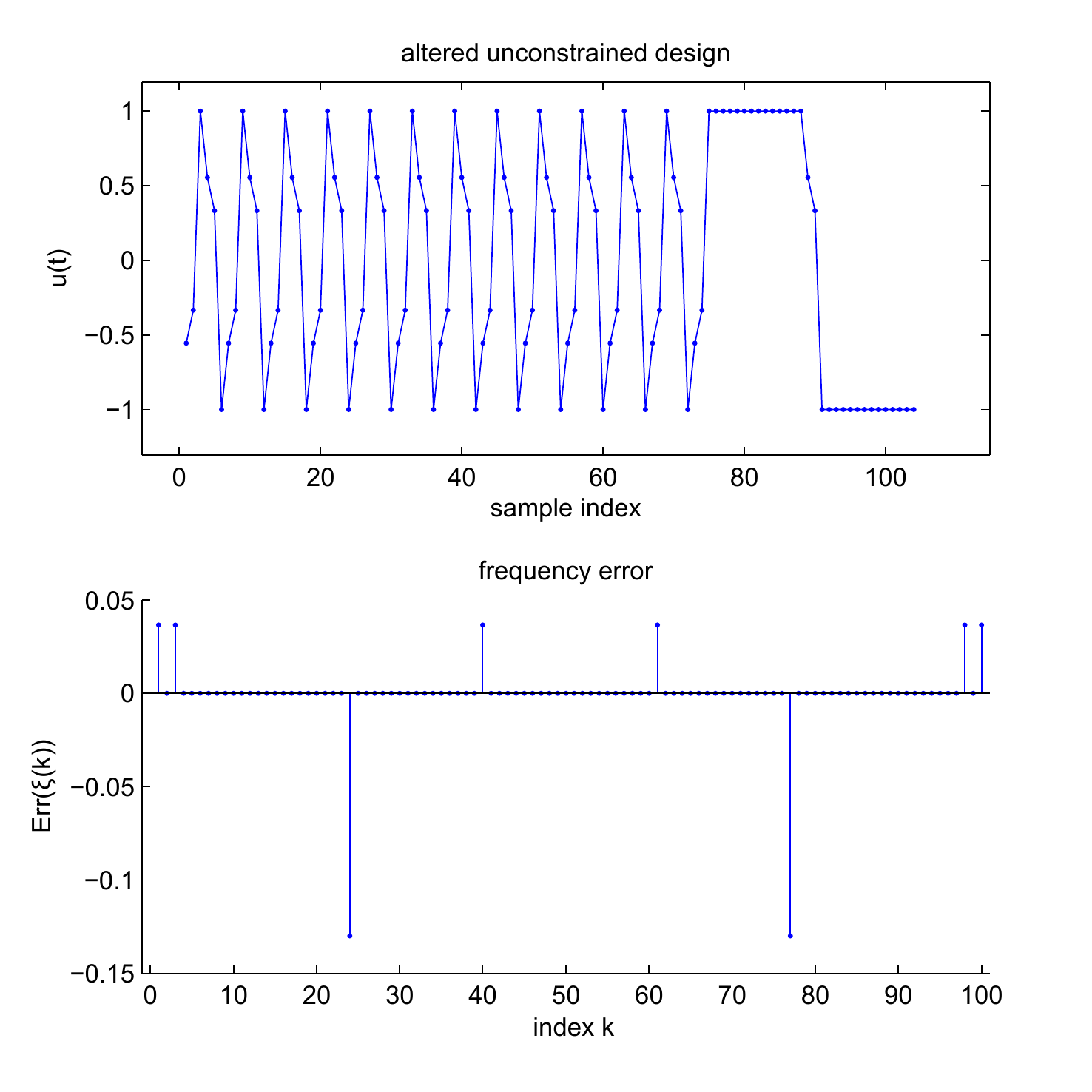}
		}}
	\caption{Input sequence  obtained from altered unconstrained design. Top: time domain signal; Bottom: difference in frequency between optimal design and time sequence for the frequency vector.}
	\label{Fig:unconsignal}
\end{figure}

Both designs will be translated into a time sequence containing 100 subsequences. During this process three approximations are considered. 
\begin{itemize}
\item After denormalization, the values of $\xi_N$ are rounded to the nearest natural number.
\item The unconstrained design  needs additional transient subsequences because condition (\ref{eq:Constraint}) is violated.
\item The constrained design  needs additional transient subsequences when the graph is disconnected.
\end{itemize}
All these approximations  lead to a decrease in the determinant of $\Fi$ and alter the total subsequence count, making the resulting time sequences suboptimal.
\begin{figure} 
	\centering{
	\resizebox{0.9\hsize}{!}{
		\includegraphics[scale=1]{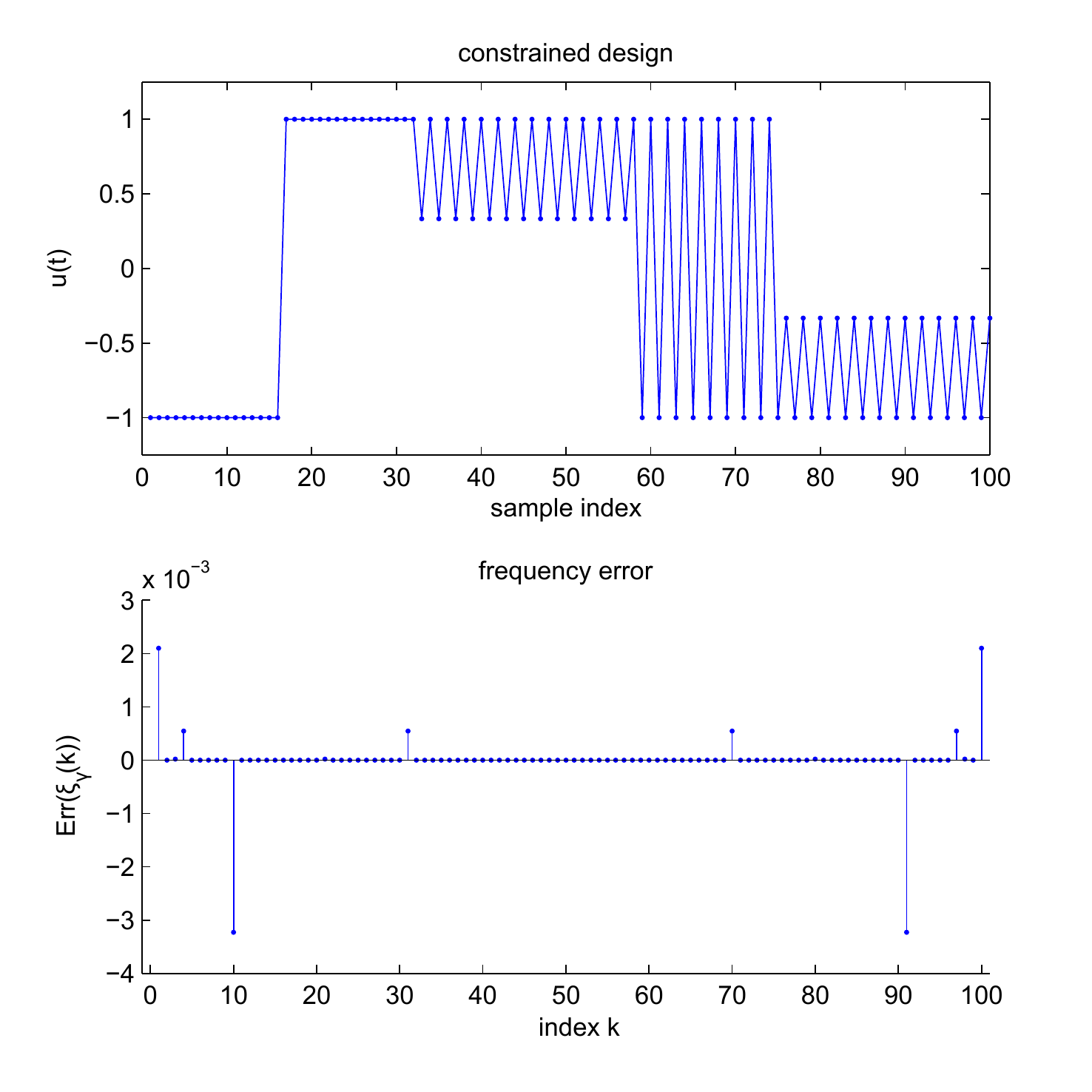}
		}}
	\caption{Input sequence with constraints. Top: time domain signal; Bottom: difference in frequency between optimal design and time sequence.}
	\label{Fig:consignal}
\end{figure}
\\First, the time sequence for the constrained design is generated. The resulting input sequence can be found in the top plot of Fig \ref{Fig:consignal}. The bottom plot depicts the difference in frequency between the optimal relative frequency vector $\xi_{opt}$ and the generated time sequence. All errors are smaller than $10^{-2}$, meaning only rounding errors are present. 
 \\ The unconstrained design does not satisfy the constraints (\ref{eq:Constraint}). Therefore, the design will be slightly altered in order to find a time sequence which realizes the unconstrained design as well as possible. The graph of the altered design can be found in Fig. \ref{Fig:modgraph}. Notice that the graph is no longer disconnected and satisfies the conditions in (\ref{eq:Constraint}). 
\\After altering the design a time sequence can be generated. The results can be found in Fig.\ref{Fig:unconsignal}. The generated sequence contains 104 samples due to roundoff errors. If a sequence of exactly 100 samples is needed some $[1,1]$ and $[-1,-1]$ subsequences could be removed. The positive frequency errors reflect the change in frequency compared to the original subsequences. The negative errors reflect the addition of the dotted arrows to the design (see Fig.\ref{Fig:modgraph}).
\subsection{Comparing Designs}
Table 3 summarizes the performance of all previous designs. In order to remove the influence of the signal length, the Fisher matrix is computed based on the relative frequencies. As a reference for comparison, the maximum determinant out of 1000 randomly generated signals is also added.
\\ Both the constrained and unconstrained designs perform two orders of magnitude better than the best randomly generated design. Out of the optimized designs, the unconstrained design has the highest determinant value. When this design is altered and realized as a time sequence, a decrease in the determinant can be observed, but the resulting sequence still outperforms the constrained design. Notice that there is no difference in determinant value between the constrained design and its corresponding time sequence.
\\ From these observations it can be concluded that it is meaningful to compute both the constrained and unconstrained design. If the unconstrained design can easily be altered to a realizable design, without too much loss in performance, it should be preferred. If not, the constrained design presents a valuable alternative, because it can always be realized.
\begin{table} 
\begin{center}
\begin{tabular}{|c|c|}
\hline 
 solution type & det(Fi)\tabularnewline
\hline 
\hline
max random signal & 9.51e+01 \tabularnewline
\hline
unconstrained design & 1.83e+03  \tabularnewline
\hline
unconstrained sequence & 1.37e+03 \tabularnewline 
\hline 
constrained design & 1.17e+03 \tabularnewline
\hline 
constrained sequence & 1.17e+03 \tabularnewline
\hline 
\end{tabular}
\end{center}
\caption{Normalized determinants of all considered designs and their corresponding time sequences}
\label{tab:Grid}
\end{table} 
\section{Computation time} \label{sec:complexity}
To evaluate the computational complexity of the dispersion-based algorithm, we compared it with the general purpose convex optimizer cvx (we choose the SeDuMi as internal solver \cite{Sturm1999,cvx}). Both algorithms are used to compute the optimal input for a given set of problems. For each problem the computation of the optimal input is performed ten times and the median of the computation time is used as measure of computing speed. 
\subsection{Stopping Condition}
In order to guarantee that the quality of the solution computed by both solvers is equal, the determinant of the Fisher information matrix of the cvx solution is used as an absolute stopping criterion for the dispersion-based algorithm. This means that the dispersion-based algorithm keeps iterating until a solution with the same or higher determinant value is found. 
\subsection{Problem Parameters}
For all problems, the system consists of an FIR filter followed by a polynomial nonlinearity. The filter coefficients correspond to the natural numbers between 1 and $n$. The nonlinearity consists of a cubic and linear term and has the same coefficients as in the numerical example. It is assumed that the cubic parameter $c_1$ is fixed during the estimation. The optimization is performed over the search space spanned by the non-overlapping symmetric set.
\\Two distinct sets of problems will be solved. For the first set of problems, we increase the number of amplitude levels between -1 and 1 while keeping the system memory constant to 2. For the second set of problems, the input set is fixed to $[-1,0,1]$ but the length of the system memory is increased.
\\ The results of these simulations are depicted in Fig. \ref{Fig:computation}. Notice that only the time for the optimization is considered. The time needed to compute the non-overlapping symmetric set, which is the same for both solvers, is depicted in Fig. \ref{Fig:computation2}.
\subsection{Effect of the Amplitude Levels}
\begin{figure} 
	\centering{
	\resizebox{1.0\hsize}{!}{
		\includegraphics[scale=1]{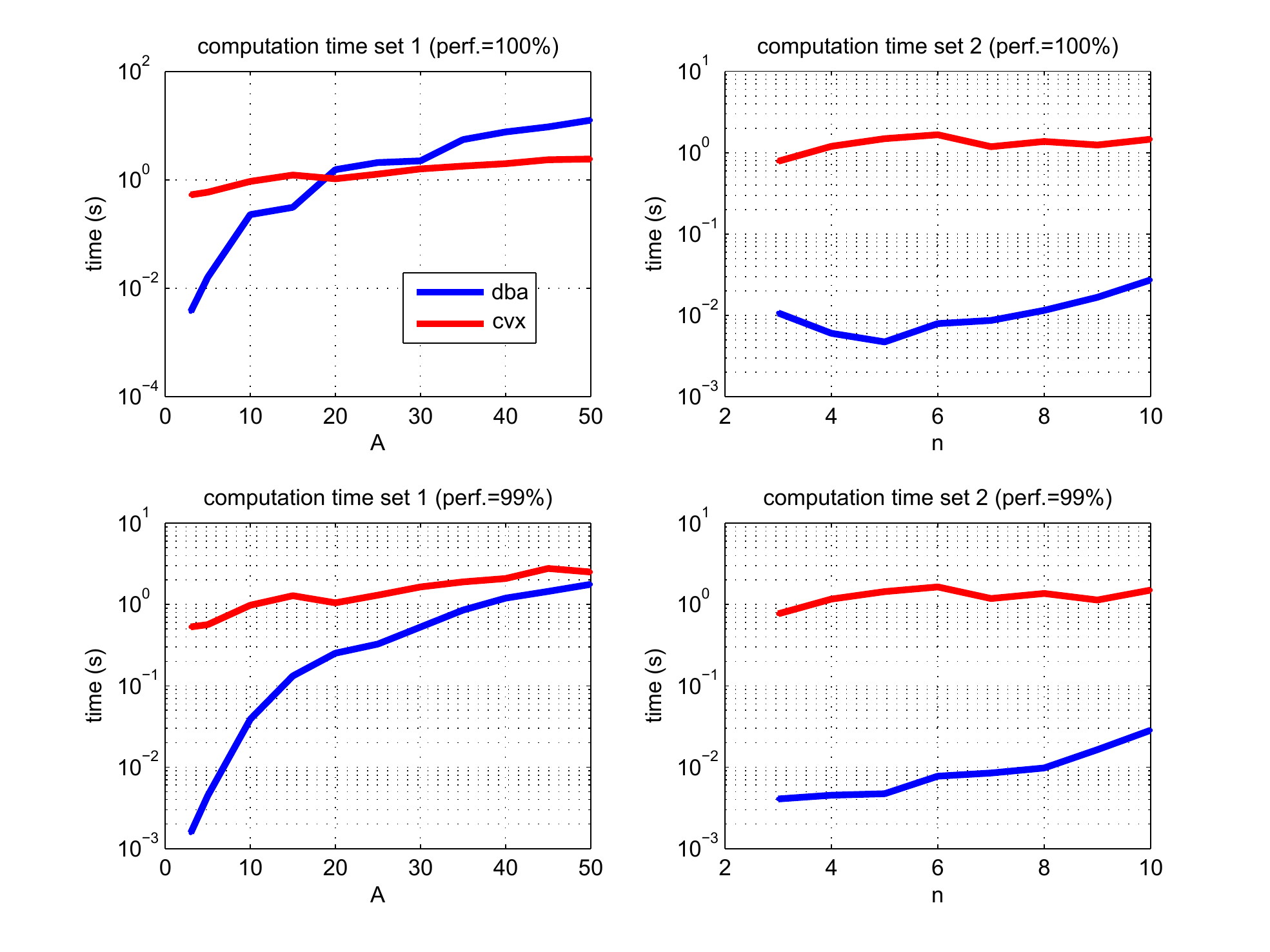}
		}}
	\caption{Median computation times obtained after 10 runs for the two problem sets. In blue the results for dispersion-based method (dba). In red the results for cvx. Only the time for the optimization process is considered.}
\label{Fig:computation}
\end{figure}

The computation times for the first set of problems are depicted in the left subplots.  We see that the computation time increases with the number of amplitude levels, regardless of the solver. This should be expected, as more amplitude levels lead to a bigger search space. For less than 20 amplitude levels the dispersion-based algorithm computes faster than cvx. Past the 20 amplitude levels the performance of the solution of cvx cannot be perfectly matched by the solution provided by dispersion-based algorithm, which leads to higher computation times. When we relax the stopping condition by allowing the dispersion-based algorithm to stop when it reaches 99\% of the determinant value found with cvx, we see that the computation times of the dispersion-based method are lower (see left, lower subplot in Fig. \ref{Fig:computation}). However, the difference between the two methods becomes smaller with increasing number of amplitude levels. 
\subsection{Effect of the Memory Length}
In the second set of problems, the input set is fixed but the length of the system memory is increased. The computation times for the second set of problems are depicted in the right subplots. We can see that the computation time increases with the length of the memory of the system. This is normal because the memory length is exponentially proportional with the dimension of the search space. More importantly, we see that the dispersion-based algorithm reaches the same performance as cvx in shorter computation times. The difference is around two orders of magnitude.
\subsection{Computing the Non-Overlapping Symmetric Set}
\begin{figure} 
	\centering{
	\resizebox{1.0\hsize}{!}{
		\includegraphics[scale=1]{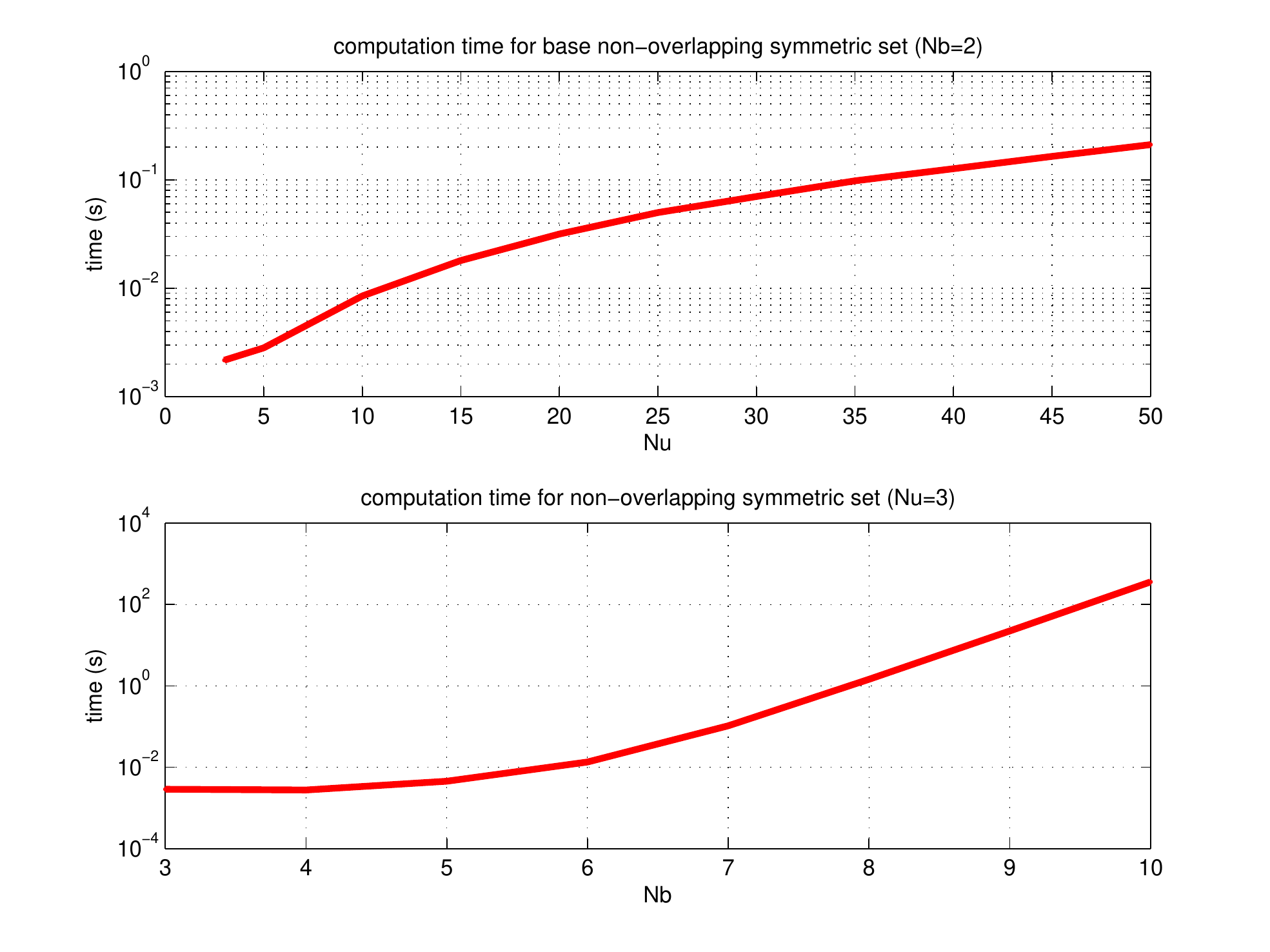}
		}}
	\caption{Median computation time for the non-overlapping symmetric set, obtained after 10 runs.}
	\label{Fig:computation2}
\end{figure}

Untill now, we only considered the time needed to perform the optimization. However, before we can perform the optimization, we need to construct the non-overlapping set and compute the Fisher information matrix for each vector in this set. In Fig. \ref{Fig:computation2} the computation time for this step is depicted for increasing problem sizes. In the top plot we see the evolution for a fixed system memory and an increasing number of amplitude levels. In the bottom plot we see the evolution for a fixed number of amplitude levels and an increasing memory length. Especially from the bottom subplot it becomes clear that computing the non-overlapping set is the most time expensive step of the OID. For a memory length of 10, computing the set takes already more than 6 minutes, while the optimization takes 2 to 3 seconds.
\subsection{Summary of the Results}
From the above results we can conclude that the dispersion-based algorithm  has a similar, if not better, performance compared to general purpose convex optimization algorithms for the presented optimal input problem. However, it turns out that the highest computational cost is not in the optimization of the design, but in the computation of the set describing the search space. As long as this bottle neck is not removed the choice of optimization algorithm is not critical. For performance comparison for large scale random optimal problems we refer to \cite{Lu2013}, where it is shown that for problems with 1000 variables or more, interior point methods have the best performance.
\section{System Memory vs Subsequence Length} \label{app}
Until now we have assumed that the length of the subsequences and the memory of the system are equal. In [28] the connection between length of the subsequences and memory length has not been examined. In this section we will illustrate why it is optimal to choose the length of the subsequences and the memory of the system equal.\\
The subsequence length determines the search space of input sequences in which we try to find the optimal sequence. Remember that this search space corresponds to a polyhedron of which the corner points can be found through the elementary cycles of the associated graph. The longer the subsequence length the more complex the graph and the more corner points the polyhedron has.\\
The memory length of the system determines how any sequence from search space of input sequences is mapped onto a frequency vector. The longer the memory length, the larger the frequency vector search space becomes. As a result less sequences will be mapped on the same frequency vector.\\
If we chose the subsequence length and memory length equal, the corner points are mapped on a set of convex independent frequency vectors. However, when the memory length is smaller than the subsequence length, this is no longer the case. Some of the frequency vectors can now be written as a convex combination of the others. In other words the effort made to compute the additional corner points is wasted since the mapping to the frequency vector space makes some corner points redundant.\\
When the memory length of the subsystem is larger than the subsequence length. All corner points will be uniquely mapped to frequency vectors. However, compared to the case of equal lengths, the search space is now smaller. This reduction may exclude more informative designs that are still realizable.
\section{Conclusion} \label{sec:Conclusion}
In this work  a solution to the problem of D-optimal input design for nonlinear FIR-type systems with an input taking a finite set of possible values has been presented.
\\By expressing the optimization problem with respect to the relative frequency vector, instead of the time sequence, the problem became convex. This convex problem was solved with an unconstrained optimization scheme based on the dispersion function. 
 \\However, it turned out that additional constraints are needed in order to guarantee that the optimal design can be realized as a time sequence. By imposing that the solution should lie in the subspace described by a symmetric and non-overlapping basis, a realizable solution was obtained that is optimal in its subspace of constrained solutions and remains numerically tractable.
\\In order to find a time sequence that realizes this optimal constrained design, the associated graph was introduced. It was shown that a path, using all edges in the graph as many times as their multiplicity indicates, corresponds to a time sequence that realizes the design. 
\\ Comparing the realization of the constrained design with the realizations of a random and unconstrained design showed that the determinant was highest for the unconstrained design. However, it can not be guaranteed that this design is realizable without a significant loss in determinant value. Therefore, the constrained design is proposed as an attractive alternative, because it can always be realized.
\\The methods presented in this paper were applied on a simple numerical example. Additionally the computational cost of the method was compared with the general purpose convex optimizer cvx. From this comparison it turned out that the dispersion based method has similar or better performance for medium sized problems.
\begin{ack}                               
This work was supported in part by the Fund for Scientific Research (FWO-Vlaanderen), by the Flemish Government (Methusalem), by the Belgian Government through the Inter university Poles of Attraction (IAP VII) Program, and the ERC Advanced Grant SNLSID.
\end{ack}
\bibliographystyle{plain}        
\bibliography{Doctoraat1bis,migrefs}  
\end{document}